\documentclass[aps,prb,twocolumn]{revtex4}

\usepackage{amsfonts}
\usepackage{subfigure}
\usepackage{amssymb}
\usepackage{amsmath}
\usepackage{bm}
\usepackage[dvips]{graphicx}
\usepackage[dvips]{color}
\usepackage{graphicx}
\usepackage{epsfig}

\usepackage{hyperref}
\usepackage[final]{changes}
\definechangesauthor[Peihao Huang]{PH}{blue}

\newcommand{\ket}[1]{\left|#1\right>}
\newcommand{\bra}[1]{\left< #1 \right|}

\newcommand{\beq}{\begin{equation}}
\newcommand{\eeq}{\end{equation}}
\newcommand{\beqa}{\begin{eqnarray}}
\newcommand{\eeqa}{\end{eqnarray}}

\begin{document}

\title{Spin relaxation in a Si quantum dot due to spin-valley mixing}

\author{Peihao Huang}
\email{peihaohu@buffalo.edu}
\author{Xuedong Hu}
\email{xhu@buffalo.edu}
\affiliation{Department of Physics, University at Buffalo, SUNY, Buffalo, NY 14260, USA}

\begin{abstract}
We study the relaxation of an electron spin qubit in a Si quantum dot due to electrical noise. In particular, we clarify how the presence of conduction-band valleys influences spin relaxation. In single-valley semiconductor quantum dots, spin relaxation is through the mixing of spin and envelope orbital states via spin-orbit interaction. In Si, the relaxation could also be through the mixing of spin and valley states. We find that the additional spin relaxation channel, via spin-valley mixing and electrical noise, is indeed important for an electron spin in a Si quantum dot. By considering both spin-valley and intra-valley spin-orbit mixings and Johnson noise in a Si device, we find that the spin relaxation rate peaks at the hot spot, where the Zeeman splitting matches the valley splitting. Furthermore, because of a weaker field dependence, the spin relaxation rate due to Johnson noise could dominate over phonon noise at low magnetic fields, which fits well with recent experiments.
\end{abstract}

\date{\today}
\maketitle

\section{Introduction}
A spin qubit is a promising candidate as an information carrier for quantum information processing,\cite{Hanson2007, Morton2011, Zwanenburg2013}
and silicon is one of the best host materials for a spin qubit.\cite{Kane1998, Morello2010, Xiao2010, Pla2012, Maune2012, Yang2013, Zwanenburg2013, Muhonen2014}  Specifically, the low abundance of isotopes with finite nuclear spins ($^{29}$Si) in natural Si significantly reduces the hyperfine interaction strength\cite{Assali2011} and the spin dephasing.\cite{Pla2012} Isotopic purification further suppresses this decoherence channel, so that Si behaves as if it is a ``semiconductor vacuum'' for a spin qubit. \cite{Muhonen2014} Spin-orbit (SO) interaction in Si is also weak because of the lighter mass of Si atoms and the lattice inversion symmetry in bulk Si.\cite{Xiao2010, Yang2013}  Therefore, as has been calculated theoretically and measured experimentally, (donor-confined) spin dephasing and relaxation times are extremely long in bulk Si. \cite{Gumann2014, Muhonen2014, Wolfowicz2013}

%
%


But Si is not perfect.  The existence of multiple conduction-band valleys\cite{Yu2010} gives additional phase factors to the electron wave function, so that interaction between donor electron spins becomes sensitively dependent on the donor positions.\cite{Koiller2001, Koiller2004, Wellard2005, Salfi2014, Gonzalez-Zalba2014}  While interface confinement and scattering can lift this degeneracy, details at the interface, whether it is surface roughness or steps, play important roles in determining the magnitude of the valley splitting $E_\mathrm{VS}$,\cite{Boykin2004, Friesen2007, Culcer2010, Friesen2010, Saraiva2010, Saraiva2011, Rahman2011, Jiang2012, Gamble2013, Dusko2014} so that device variability is large.  Experimentally measured $E_\mathrm{VS}$ ranges from vanishingly small, to several hundreds of $\mu$eV,\cite{Shaji2008, Yang2013, Hao2014} to possibly a few meV.\cite{Takashina2006} Furthermore, to achieve controllability, spin qubits are generally located near or at the interface between the host and the barrier materials.  Dangling bonds, charge traps, and other defects are inevitably present at the many interfaces of a semiconductor heterostructure, and the coherence properties of a spin qubit in a nanostructure are not as clearly understood and measured as in bulk Si.

With pure dephasing strongly suppressed in Si, spin relaxation becomes an important indicator of decoherence for a spin qubit.  Spin relaxation could come directly from magnetic noise in the environment, or from electrical noise via spin-orbit or exchange interaction.  Indeed, for a single spin in a quantum dot, we have shown\cite{Huang2014} that electrical noise from the circuits or surrounding traps could be an important cause for spin relaxation, particularly at a smaller qubit energy splitting.  In this previous study, however, we only considered intra-valley orbital dynamics for an electron in Si. On the other hand,  it has been shown experimentally and theoretically that the presence of valleys in Si can significantly modify spin relaxation through spin-valley mixing, and a relaxation hot spot appears at the degeneracy point where the Zeeman splitting matches the valley splitting.\cite{Yang2013,Tahan2014}

In this paper, we study spin relaxation of a single QD-confined electron in Si due to the presence of electrical noises (including Johnson noise, phonon noise, and the $1/f$ charge noise). \added[]{One relaxation mechanism involves the mixing of spin and valley states, which should be particularly important when Zeeman energy $E_Z$ is comparable with valley splitting $E_\mathrm{VS}$.  Another mechanism involves the mixing of spin and orbital states within one conduction band valley, which is important at high magnetic fields. By considering both of these mechanisms and various electrical noises, such as phonon noise and Johnson noise, we find that the spin-valley mixing is indeed an important spin relaxation channel for an electron spin in a Si quantum dot.  We also find that, because of a weaker field-dependence, spin relaxation due to Johnson noise through the mixing of spin and valley states could dominate over phonon noise and intra-valley scattering (relaxation due to mixing of spin and higher orbital states) at low magnetic fields.  Our numerical results fit quite well with recent experimental measurements.}


The rest of the paper is organized as follows. In Sec.~\ref{sec:system} we set up the system Hamiltonian and describe the mechanism of spin-valley mixing. In Sec.~\ref{sec:spinrelax} we derive explicitly the spin relaxation rate due to spin-valley mixing and electrical noise. In Sec.~\ref{sec:results} we evaluate the spin relaxation rates due to Johnson and phonon noises, and we compare the different spin relaxation mechanisms. Finally, conclusions are drawn in Sec.~\ref{sec:conclusion}. In the Appendices we discuss the field dependence of the spin relaxation, the effects of $1/f$ noise, and the phonon noise spectrum in more detail.


\section{System Hamiltonian}
\label{sec:system}

We consider an electron in a gate-defined quantum dot in a Si heterostructure (whether a Si/SiO$_x$ or a Si/SiGe structure). The growth-direction ([001]-direction in this paper) confinement is taken to be very strong, so that we focus on the in-plane dynamics of the confined electron.  The strong field and strain at the interface lower the degeneracy of the Si conduction band by raising the energy of four of the valleys relative to the other two (in this case $z$ and $-z$ valleys).  Moreover, scattering off the smooth interface further mixes and splits the two low-energy valleys.  We label the two valleys as $+$ and $-$, with valley splitting $E_\mathrm{VS}$.  At this smooth-interface limit, and without considering the spin-orbit interaction, the valley degrees of freedom and the intra-valley effective-mass dynamics can be separated, so that the electron wave function can be written as $|v, i, \alpha \rangle$, where $v = \pm$ is the index for the two lowest-energy eigen-valleys, $i$ is the orbital excitation index within an eigen-valley, and $\alpha = \uparrow$ or $\downarrow$ is the spin index.


\added[]{In the following, we first consider explicitly spin relaxation due to spin-valley mixing, which is important when $E_Z \sim E_\mathrm{VS}$. Later, in Sec.III, we compare these results with spin relaxation due to intra-valley spin-orbital mixing, which is a spin relaxation mechanism that is well known in the literature. By considering various electrical noises, we can then identify the dominant spin relaxation mechanism in different regimes.}

We consider a quantum dot for which the lateral confinement is sufficiently strong ($> 1$ meV), so that the intra-valley orbital level spacing is larger than the valley splitting $E_\mathrm{VS}$ (which is up to a fraction of 1 meV in general). In this case, we can focus on the effects of spin-valley mixing, and we neglect the intra-valley excitation, \added[]{particularly when the Zeeman energy is close to the valley splitting, and is much less than intra-valley orbital level spacing}. In this limit, only the lowest four spin-valley states are relevant, all having the intra-valley ground orbital state.  These four states (with an implicit common orbital index $i = 0$) are denoted as $\ket{1} = \ket{-, \downarrow}$, $\ket{2}=\ket{-,\uparrow}$, $\ket{3}=\ket{+,\downarrow}$, and $\ket{4}=\ket{+,\uparrow}$.
Within the space spanned by these four lowest-energy spin-valley product states, the total Hamiltonian for the QD-confined electron is given by
\beqa
H&=& H_0 + H_\mathrm{SV} + H_\mathrm{noise}, \\
H_0 & = & \sum_i\frac{\epsilon_i}{2}\ket{i}\bra{i}, \nonumber \\
H_\mathrm{SV} & = & \frac{\Delta_{23}}{2}\ket{2}\bra{3}+\frac{\Delta_{14}}{2}\ket{1}\bra{4}, \nonumber\\
H_\mathrm{noise} & = & -e\vec{E}\cdot \left[\vec{r}^{--}\sum_{i=1,2}\ket{i}\bra{i} + \vec{r}^{++}\sum_{i=3,4}\ket{i}\bra{i}\right],\nonumber\\
&& -e\vec{E}\cdot \vec{r}^{-+}(\ket{1}\bra{3}+\ket{2}\bra{4}) + \mathrm{H.c.} \nonumber
\eeqa
Here $H_0$ contains valley and Zeeman splitting, with $\epsilon_i/2$ being the energies of the product states in the absence of SO interaction and the environmental noise.  Specifically, $\epsilon_4=-\epsilon_1=E_\mathrm{VS}+g\mu_BB$, $\epsilon_3=-\epsilon_2=E_\mathrm{VS}-g\mu_BB$.  $H_\mathrm{SV}$ represents spin-valley (SV) mixing due to the SO interaction, with $\Delta_{23}$ and $\Delta_{14}$ the SV mixing energy: $\Delta_{23} = 2\bra{2} H_{SO} \ket{3} = 2\bra{-,\uparrow} H_{SO} \ket{+,\downarrow}$ and $\Delta_{14} = 2\bra{1} H_{SO} \ket{4} = 2 \bra{-,\downarrow} H_{SO} \ket{+,\uparrow}$. Here the SO interaction is $H_{SO} = \alpha_{-}p _{y}\sigma _{x}+\alpha_{+}p _{x}\sigma _{y}$, with the interaction strength $\alpha_{\pm }\equiv \left( \alpha_D \pm \alpha_R \right)$, and the $x$ and $y$ axes along the [$110$] and [$\bar{1}10$] directions (which also define the plane of the quasi-2D quantum dot).\cite{Golovach2004, Huang2013}  Here $\alpha_D$ and $\alpha_R$ are the Dresselhaus and Rashba SO interaction constants. The Dresselhaus SO interaction arises from the bulk inversion asymmetry, which in a Si QD could be from the interface disorder, while Rashba SO interaction arises from the structure inversion asymmetry and is tunable through the electric field across the QD. Lastly, $H_\mathrm{noise}$ contains the electrical noise from the environment, with $E(\vec{r})$ the noise electric field. It could come from Johnson noise, $1/f$ charge noise, phonon noise, etc..  Here $\vec{r}^{-+}=\bra{-}\vec{r}\ket{+}$ is the electric dipole matrix element between different valley states, which could arise from disorders at the interfaces of the QD.\cite{Gamble2013}



\begin{figure}[t]
  \includegraphics[width=0.4\textwidth]{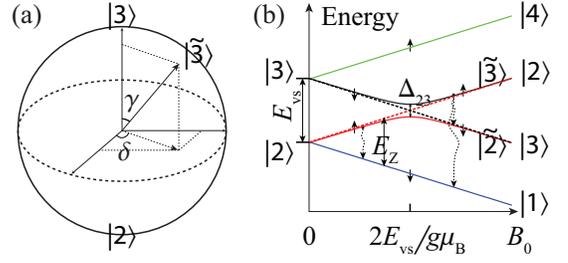}\\
  \caption{(a) The relations between the product states $\ket{3}$ (or $\ket{2}$) and the eigenstate $\widetilde{\ket{3}}$; $\gamma$ and $\delta$ are the polar and azimuthal angles of the orientation of the eigenstate $\widetilde{\ket{3}}$ in the basis of product states. (b) The level diagram of the system as a function of the applied magnetic field. States $\ket{1}$, $\ket{2}$, $\ket{3}$, and $\ket{4}$ are the product states, and states $\widetilde{\ket{2}}$ and $\widetilde{\ket{3}}$ are the eigenstates after the SV mixing. $E_\mathrm{VS}$ and $E_Z$ are the valley splitting and Zeeman splitting, respectively. The small arrows on the energy levels indicate the spin orientations.}\label{scheme}
\end{figure}

We first find the eigenstates of the confined electron in the presence of spin-valley mixing but without environmental noises. As indicated in Fig.~\ref{scheme}, states $|1\rangle$ and $|4\rangle$ are always well separated energetically, by both the valley and the Zeeman splitting, so that we neglect the mixing of states $|1\rangle$ and $|4\rangle$ by $\Delta_{14}$ in this study.  On the other hand, near $g\mu_B B = E_\mathrm{VS}$, states $\ket{2}$ and $\ket{3}$ are strongly mixed by the spin-valley coupling $\Delta_{23}$.  This degeneracy point is called a spin relaxation hot spot. \cite{Fabian1998, Stano2006}  $\Delta_{23}$ is in general a complex number, and it can be written as $\Delta_{23}=\Delta_1+i\Delta_2$ and $\Delta=|\Delta_{23}|$.  The eigenstates for $H_0 + H_\mathrm{SV}$ are thus \{$\ket{1}$, $\widetilde{\ket{2}}$, $\widetilde{\ket{3}}$, $\ket{4}$\}, where
\beqa
\widetilde{\ket{3}}&=&\cos(\gamma/2)e^{-i\delta/2}\ket{3}+\sin(\gamma/2)e^{i\delta/2}\ket{2},\\
\widetilde{\ket{2}}&=&-\sin(\gamma/2)e^{-i\delta/2}\ket{3}+\cos(\gamma/2)e^{i\delta/2}\ket{2} \,.
\eeqa
Here $\gamma=\arctan(|\Delta|/\epsilon_3)$ and $\delta=\arctan(\Delta_2/\Delta_1)$.  The energy splitting between $\widetilde{\ket{3}}$ and $\widetilde{\ket{2}}$ is $\widetilde{\epsilon}_3=\sqrt{\epsilon_3^2+\Delta^2}$. When the magnetic field is along the [110] axis as in Ref. \onlinecite{Yang2013}, the spin-valley mixing matrix element $\Delta_{23}$ can be expressed as (see Appendix \ref{append:B-orientation}) \cite{Tahan2014}
\beq
\Delta_{23} = 2 m^* E_\mathrm{VS} \alpha_+ r _{x}^{-+}/\hbar.
\eeq
where the relationship $\vec{p}^{-+}=\bra{-}\vec{p}\ket{+}=im^*E_\mathrm{VS} \vec{r}^{-+}/\hbar$ has been employed.

\section{Spin relaxation}
\label{sec:spinrelax}

\subsection{Spin relaxation due to spin-valley mixing}


With states $\widetilde{\ket{2}}$ and $\widetilde{\ket{3}}$ being spin-valley mixed, and assuming that the electric dipole matrix element between the two eigen-valleys is non-vanishing, any electrical noise, which couples states with the same spin orientation, can induce transitions between them and from them to the other two eigenstates.  The transition rate is proportional to the amount of spin-valley mixing, and to the spectrum of the noisy electric field $E(\vec{r})=\vec{\nabla}U_\mathrm{noise}(\vec{r})/e$, where $U_\mathrm{noise}(\vec{r})$ captures the electrical potential of the noise in the system, such as Johnson noise, $1/f$ charge noise or phonon noise, which will be discussed later.

Experimentally, in the preparation of a spin-up initial state, the electron orbital and valley states are kept in the lowest eigenstates in order to avoid the unnecessary mixing of the spin and orbital dynamics.  The most relevant spin relaxation processes involve the relaxation of either state $\widetilde{\ket{2}}$ or state $\widetilde{\ket{3}}$ because of the experimental difficulty in making measurements close to the spin-valley crossing point, the small magnitude of spin-valley mixing ($\Delta\sim 10$ neV), and the energy-selective nature of resonant tunneling \cite{Yang2013}.  Specifically, we consider the following situations: when $E_Z < E_\mathrm{VS}$, a spin-up electron is loaded only into the energy eigenstate $\widetilde{\ket{2}}$; while when $E_Z > E_\mathrm{VS}$, it is only loaded into the energy eigenstate $\widetilde{\ket{3}}$.



In the low-field regime when $E_Z<E_\mathrm{VS}$, spin relaxation occurs from state $\widetilde{\ket{2}}$ to the ground state $\ket{1}$. The spin relaxation rate is \cite{Huang2014}
\beq
\Gamma_{\widetilde{2}1} = \frac{2e^2}{\hbar^2}\int_{-\infty}^{\infty} \overline{\bra{1}\vec{E}\cdot \vec{r}\widetilde{\ket{2}}\widetilde{\bra{2}}\vec{E}(t)\cdot \vec{r}\ket{1}}\cos(\Delta E_{\widetilde{2}1} t) dt
\eeq
where $\Delta E_{\widetilde{2}1}$ is the energy difference between state $\widetilde{\ket{2}}$ and $\ket{1}$, and $\overline{x}$ means an average of $x$ with respect to the noise electric field.  In the case of quantum noise, this should be an ensemble average.  Separating the noise electric field from the coupling matrix element, the spin relaxation rate can also be expressed as
\beq
\Gamma_{\widetilde{2}1} = \frac{4\pi e^2}{\hbar^2} \sum_i\left|\bra{1}{r}_i\widetilde{\ket{2}}\right|^2S_{ii}^E(\Delta E_{\widetilde{2}1}),
\eeq
where $S_{ii}^E(\omega)\equiv\frac{1}{2\pi}\int_{-\infty }^{+\infty }d\tau \overline{E_{i}(0)E_{i}(\tau)} \cos(\omega\tau)$ is the noise spectrum ($i=x,y,z$), and we have assumed that noise in different directions are not correlated. The relevant transition matrix element in this case is $\bra{1}\vec{r}\widetilde{\ket{2}}=-\vec{r}^{-+}\sin(\gamma/2)$, which is proportional to the transition matrix elements $\vec{r}^{-+}$ between the $\pm$ valleys.


In the high-field regime when $E_Z>E_\mathrm{VS}$, state $\widetilde{\ket{3}}$ is loaded initially.  The electron can then relax to both $\widetilde{\ket{2}}$ and $|1\rangle$ due to spin-valley mixing and inter-valley transitions.  Both these processes involve an apparent spin flip.  The relaxation rates are
\beqa
\Gamma_{\widetilde{3}1} &=& \frac{4\pi e^2}{\hbar^2}\sum_i\left|\bra{1}{r}_i\widetilde{\ket{3}}\right|^2S_{ii}^E \left(\Delta E_{\widetilde{3}1}\right), \\
\Gamma_{\widetilde{3}\widetilde{2}} &=& \frac{4\pi e^2}{\hbar^2}\sum_i\left|\widetilde{\bra{2}}{r}_i\widetilde{\ket{3}}\right|^2 S_{ii}^E \left(\Delta E_{\widetilde{3}\widetilde{2}} \right),
\eeqa
where the relevant matrix elements are $\bra{1}{r}_i\widetilde{\ket{3}}={r}_i^{-+}\cos(\gamma/2)$, $\widetilde{\bra{2}} {r}_i \widetilde{\ket{3}} = \left(r^{--}_i - r^{++}_i \right) (\sin\gamma)/2 $.  Below we will focus on the spin-valley transition from $\widetilde{\ket{3}}$ to $|1\rangle$ when $E_Z>E_\mathrm{VS}$.

Since the SO mixing element is much less than Zeeman energy, $\Delta\ll E_Z$, spin-valley relaxation rates $\Gamma_{\widetilde{3}1}$ and $\Gamma_{\widetilde{2}1}$ take the same algebraic form, \added[]{and the energy transfer involved, $\Delta E_{\widetilde{2}1}$ and $\Delta E_{\widetilde{3}1}$, can both be approximated by $\hbar\omega_Z$ in their respective field regime.} We thus use $\Gamma_\mathrm{SV}$ to denote spin relaxation rate due to SV mixing \added[]{(in the next subsection we will discuss the spin relaxation rate $\Gamma_\mathrm{SO}$ due to intra-valley SO mixing, which involves higher electron orbital states but within the same valley), so that $\Gamma_\mathrm{SV}=\Gamma_{\widetilde{2}1}$ when $E_Z<E_\mathrm{VS}$; and $\Gamma_\mathrm{SV}=\Gamma_{\widetilde{3}1}$ when $E_Z>E_\mathrm{VS}$. The resulting spin relaxation rate is},
\beqa
&&\Gamma_\mathrm{SV}=\frac{2\pi e^2}{\hbar^2}\sum_i\left|r^{-+}_i\right|^2 S_{ii}^{E}(\omega_Z)F_\mathrm{SV}(\omega_Z),\label{Gamma_SV}\\
&&F_\mathrm{SV}(\omega_Z)=1 - \left[ 1 + \frac{\Delta^2}{(E_\mathrm{VS}-\hbar\omega_Z)^2} \right]^{-\frac{1}{2}}, \label{FSV}
\eeqa
where $F_\mathrm{SV}(\omega_Z)$ is from the dipole matrix elements such as $|\bra{1}r\widetilde{\ket{2}}|^2=|r_i^{-+}|^2 |\sin(\gamma/2)|^2=|r_i^{-+}|^2 F_\mathrm{SV}/2$ when $E_Z<E_\mathrm{VS}$. In other words, spin relaxation is now allowed because $r^{-+}$ allows inter-valley charge transitions, while $F_\mathrm{SV}$ allows spin and valley-charge states to mix. More specifically, $F_\mathrm{SV}$ contains the field dependence of the spin-valley mixing. Its $\omega_Z$ dependence comes directly from the applied field. As shown in Eq.~\ref{FSV}, $F_\mathrm{SV}(\omega_Z)$ peaks at the degeneracy point $\hbar\omega_Z=E_\mathrm{VS}$, where $F_\mathrm{SV}=1$ and has a width of $2 \Delta$ because of the maximum mixing of the valley states at the degeneracy point. Away from it, when $|E_\mathrm{VS}-\hbar\omega_Z|\gg \Delta$,
\begin{eqnarray}
F_\mathrm{SV}(\omega_Z) & \approx & 1 - \left[ 1 - \frac{\Delta^2}{2 (E_\mathrm{VS}-\hbar\omega_Z)^2} \right] \nonumber \\
& \approx & \frac{\Delta^2}{2 (E_\mathrm{VS}-\hbar\omega_Z)^2}.
\end{eqnarray}

On the low-energy side of the peak, with $\hbar\omega_Z \ll E_\mathrm{VS}$, $F_\mathrm{SV} \sim \Delta^2/2E_\mathrm{VS}^2 \ll 1$ approaches a small constant that is $\sim 0$; On the high-energy side, with $\hbar\omega_Z \gg E_\mathrm{VS}$, $F_\mathrm{SV} \sim \Delta^2/2\omega_Z^2$, which again approaches 0 as $\omega_Z$ increases. This clear peak structure means that the spin-valley mixing induced spin relaxation is the most significant near the degeneracy point between $|2\rangle$ and $|3\rangle$.

In the cases when $r_i^{--}=r_i^{++}$ for the transition matrix elements between the valley states,\cite{Yang2013} which implies that valley energy shift due to the electrical noise is the same in both valleys, the relaxation rate $\Gamma_{\widetilde{3}\widetilde{2}}$ vanishes.  $\Gamma_\mathrm{SV}$ is then the only spin relaxation channel due to spin-valley mixing.


For the sake of completeness, we now consider the relaxation of state $\ket{4}$. The relaxation of state $\ket{4}$ has two possible origins: The first is the relaxation to $\widetilde{\ket{2}}$ and $\widetilde{\ket{3}}$.  This is valley relaxation due to electrical noise, with a relaxation rate that is proportional to $|\bra{4}{r}_i\widetilde{\ket{2}}|^2 + |\bra{4}{r}_i\widetilde{\ket{3}}|^2=|{r}_i^{-+}|^2$, so that $\Gamma_{4\widetilde{2}}+\Gamma_{4\widetilde{3}}=\frac{2\pi e^2}{\hbar^2}\sum_i\left|r^{-+}_i\right|^2 S_{ii}^{E}(\omega_Z)$. The spin-valley relaxation of state $\ket{4}$ to $\widetilde{\ket{2}}$ is identical to relaxation from $\widetilde{\ket{3}}$ to $\ket{1}$, because the transition matrix elements, the degree of spin-valley mixing, and the energy splitting are all the same for these two transitions.  The second relaxation mechanism for state $\ket{4}$ is the relaxation due to spin-valley mixing of $\ket{4}$ and $\ket{1}$, which has been omitted at the beginning, since the effect of $\ket{4}-\ket{1}$ mixing is suppressed by the large energy separating $\ket{4}$ and $\ket{1}$.  However, when considering relaxation of state $\ket{4}$, this particular spin valley mixing could certainly lead to additional relaxation. In the following, we focus on the spin relaxation $\Gamma_\mathrm{SV}$ of states $\widetilde{\ket{2}}$ and $\widetilde{\ket{3}}$ with the flipping of spin-up state to spin-down state.

The spin relaxation mechanism discussed here is a consequence of spin-valley mixing and finite electric dipole matrix elements between the valley states. Therefore, as shown in the Eq.~(\ref{Gamma_SV}), the relaxation rate $\Gamma_\mathrm{SV}$ is proportional to the matrix elements $\left|r^{-+}_i\right|^2$ and the function $F_\mathrm{SV}(\omega_Z)$, which captures the extent of SV mixing. Finally, the $\omega_Z$ dependence of $\Gamma_\mathrm{SV}$ is given by $S_{ii}^{E}(\omega_Z)F_\mathrm{SV}(\omega_Z)$, which depends on the specific noise spectrum $S_{ii}^E(\omega)$.

\subsection{Spin relaxation due to intra-valley SO mixing}

\added[]{Spin relaxation due to spin-valley mixing is particularly important when $E_Z$ is comparable with $E_\mathrm{VS}$ and is much less than orbital level spacing $\hbar \omega_d$.  As B-field increases, higher-energy orbital states also start to contribute to spin relaxation significantly.} For comparison, we include in our discussion below spin relaxation due to intra-valley SO mixing \added[]{(higher energy p-orbitals are involved)}, which has been studied extensively in the literature, especially for spin qubit in GaAs QD. \cite{Khaetskii2001, Erlingsson2002, Golovach2004, Tahan2005, Marquardt2005, SanJose2006, Trif2008, Yang2013, Tahan2014, Huang2014, Jing2014} For spin qubit in Si QD, this intra-valley SO mixing induced spin relaxation is also present, and is important in high B-field due to the stronger B-field dependence. \cite{Tahan2014, Huang2014} \added[]{We use the existing results in the literature, and} the corresponding spin relaxation rate is \cite{Golovach2004, Tahan2014, Huang2014}
\begin{eqnarray}
\Gamma_\mathrm{SO} &=& \frac{4\pi e^2}{\hbar^2}\frac{\omega _{Z}^2}{\omega _{d}^{4}} S_{xx}^E(\omega_Z)F_\mathrm{SO}(\theta,\varphi) ,\label{1T1}
\end{eqnarray}
where $\omega_d$ is the lateral confinement strength of QD, $\omega_Z$ is the Zeeman frequency, and $S_{xx}^E(\omega)$ is the Fourier spectrum of the correlation of in-plane electric field fluctuations (in-plane electrical noise is assumed to be isotropic, and out of plane electrical noise is neglected because of the strong vertical confinement at the interface). $F_\mathrm{SO}$ contains the dependence on the SO interaction strength and the orientation of magnetic field. For a magnetic field along [110] direction as in Ref.~\onlinecite{Yang2013}, we have $F_\mathrm{SO}=\alpha_+^2$. \cite{Huang2014}

\added[]{In a general calculation of spin relaxation in a Si QD, both spin relaxation mechanisms, namely relaxation due to spin-valley mixing ($\Gamma_\mathrm{SV}$) and relaxation due to intra-valley SO mixing ($\Gamma_\mathrm{SO}$), need to be accounted for.  We consider both in our calculations below in order to achieve a comprehensive understanding of spin relaxation.}

\section{Results}
\label{sec:results}



In this section, we present spin relaxation rates for different noises, and we compare the spin relaxation channels due to SV mixing and intra-valley SO mixing.  We mainly focus on the electrical noise from Johnson noise and phonon noise.  Although $1/f$ charge noise is ubiquitous as well in semiconductor material, we do find that spin relaxation due to $1/f$ noise is much slower compared to that due to Johnson and phonon noise.  Thus we only give a brief discussion on $1/f$ noise in Appendix \ref{append::1fnoise}.


\subsection{Johnson noise}
\label{Johnson}

\begin{figure}[t]
  \includegraphics[width=0.4\textwidth]{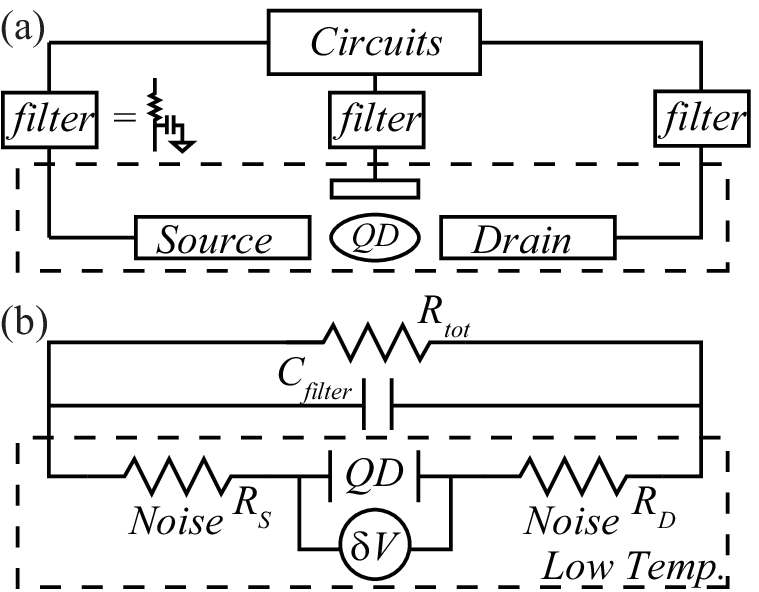}\\
  \caption{Schematic diagram of a gate-defined QD and a source of Johnson noise. (a) Schematic diagram of a gate-defined QD. The device inside the dilution refrigerator is in the dashed line box. The high-temperature Johnson noise is normally filtered in the spin qubit experiments. Only the Johnson noise from circuits inside the dilution refrigerator causes strong voltage fluctuations, and induces spin relaxation. (b) Simplified circuits diagram, where only Johnson noise from the resistances of source and drain are relevant.}\label{circuits}
\end{figure}

Johnson noise is the electromagnetic fluctuations in an electrical circuit. For a gate-defined QD, Johnson noise inside the metallic gates, such as the source and drain circuits, could give rise to strong electrical fluctuations acting on the QD, and it could induce spin decoherence for the electron confined in the QD.

The spectrum of Johnson noise is given by \cite{Weiss1999}
\begin{equation}
S_{V}\left( \omega \right) =2\xi \omega \hbar ^{2}f_c(\omega_Z)\coth \left( \hbar \omega /2k_{B}T\right), \label{Sv_John}
\end{equation}
where $S_V$ is the spectrum of electrical voltage $S_{V}(\omega)=\frac{1}{2\pi }\int_{-\infty }^{+\infty }\overline{V\left( 0\right)  V\left( t\right) } \cos \left( \omega t\right) dt$, $\xi =R/R_{k}$ is a dimensionless constant, $R_{k}=h/e^{2}=26$ k$\Omega $ is the quantum resistance, and $R$ is the resistance of the circuit. $f_c(\omega)=1/[1+\left( \omega/\omega _{R}\right) ^{2}]$ is a natural cutoff function for Johnson noise, where $\omega _{R}=1/RC$ is the cutoff frequency, and $C$ is capacitors in parallel with the resistance $R$.

As shown in Fig.~\ref{circuits}, the Johnson noise of the circuits outside the dilution refrigerator is generally well-filtered. Thus we consider only Johnson noise of the low-temperature circuit inside a dilution refrigerator. The corresponding spectrum of electric field is $S_{ii}^{E} \left( \omega \right) = S_{V} (\omega)/(el_{0})^{2}$, where $l_{0}$ is the length scale between the source and drain.  Accordingly, the spin relaxation rate due to SV mixing and Johnson noise is
\begin{equation}
\Gamma_\mathrm{SV}=\frac{2\pi}{\hbar^2} S_{V}(\omega_Z)F_\mathrm{SV}(\omega_Z)\sum_i\left|r^{-+}_i\right|^2/l_{0}^{2},\label{1T1_Johnson}
\end{equation}
where $F_\mathrm{SV}(\omega_Z)$ is given by Eq.~(\ref{FSV}).  The small capacitance of source and drain leads means that the cutoff frequency $\omega_R$ satisfies $\omega_R\gg\omega_Z$, so that the cutoff function $f_c(\omega_Z)\approx 1$.  The low temperature environment ensures $\coth \left( \hbar \omega /2k_{B}T\right)\approx 1$.  Therefore, the $\omega_Z$ dependence of $\Gamma_\mathrm{SV}$ is determined by $\omega_Z F_\mathrm{SV}(\omega_Z)$.

Compared with the intra-valley SO mixing mechanism, where $\Gamma_\mathrm{SO}$ shows an $\omega_Z^3$ dependence,\cite{Huang2014} $\Gamma_\mathrm{SV}$ is linearly dependent on $\omega_Z$ at low fields, when $|E_\mathrm{VS} \gg \hbar\omega_Z|$ so that $F_\mathrm{SV} (\omega_Z) \sim \Delta^2/2E_\mathrm{VS}^2$. Because of this weaker field dependence, the spin relaxation rate $\Gamma_\mathrm{SV}$ would dominate over $\Gamma_\mathrm{SO}$ at very low magnetic fields.  On the other hand, at high fields, when $\hbar\omega_Z \gg E_\mathrm{VS}$, we have $F_\mathrm{SV} (\omega_Z) \sim \Delta^2/2\omega_Z^2$, then $\Gamma_\mathrm{SV} \propto 1/\omega_Z$: the relaxation rate is slower as the external field increases.  Thus, at high fields the intra-valley spin relaxation should dominate over inter-valley spin relaxation.

Below we carry out numerical calculations of the spin relaxation rate in a small Si/SiO$_2$ QD.  Based on the parameters of Ref.~\onlinecite{Yang2013}, the valley splitting here is set as $E_\mathrm{VS}=0.33$ meV, the dot confinement energy is $\hbar \omega_{d}=8$ meV, and the electric dipole matrix elements for the valley states are set as $r_i^{--}=r_i^{++}=0$ and $r_x^{-+}=r_y^{-+}=r_z^{-+}=1.1$ nm.  The magnetic field is along the $[110]$ direction, and the SO interaction strength for Si is set as $\alpha_R=45$ m/s and $\alpha_D=0$ m/s.\cite{Wilamowski2002, Tahan2005, Prada2011, Yang2013} We use the bulk g-factor $g=2$, and in the lowest two valleys the electron effective mass is $m^{\ast}=0.19m_{0}$, where $m_{0}$ is the free electron rest mass. For Johnson noise parameters, we choose the resistance $R=2$ k$\Omega$, length scale $l_{0}=100$ nm and temperature $T=0.15$ K.  The magnitude of the chosen resistance allows us to obtain the best numerical fit to the experimental data (with the rest of the parameters chosen according to Ref.~\onlinecite{Yang2013}).  While resistances of the thin metallic gates at low temperatures are generally much smaller than 2 k$\Omega$, the resistance of other elements such as 2DEG channels can easily be in this order.


\begin{figure}[tb]
\includegraphics[scale=0.44]{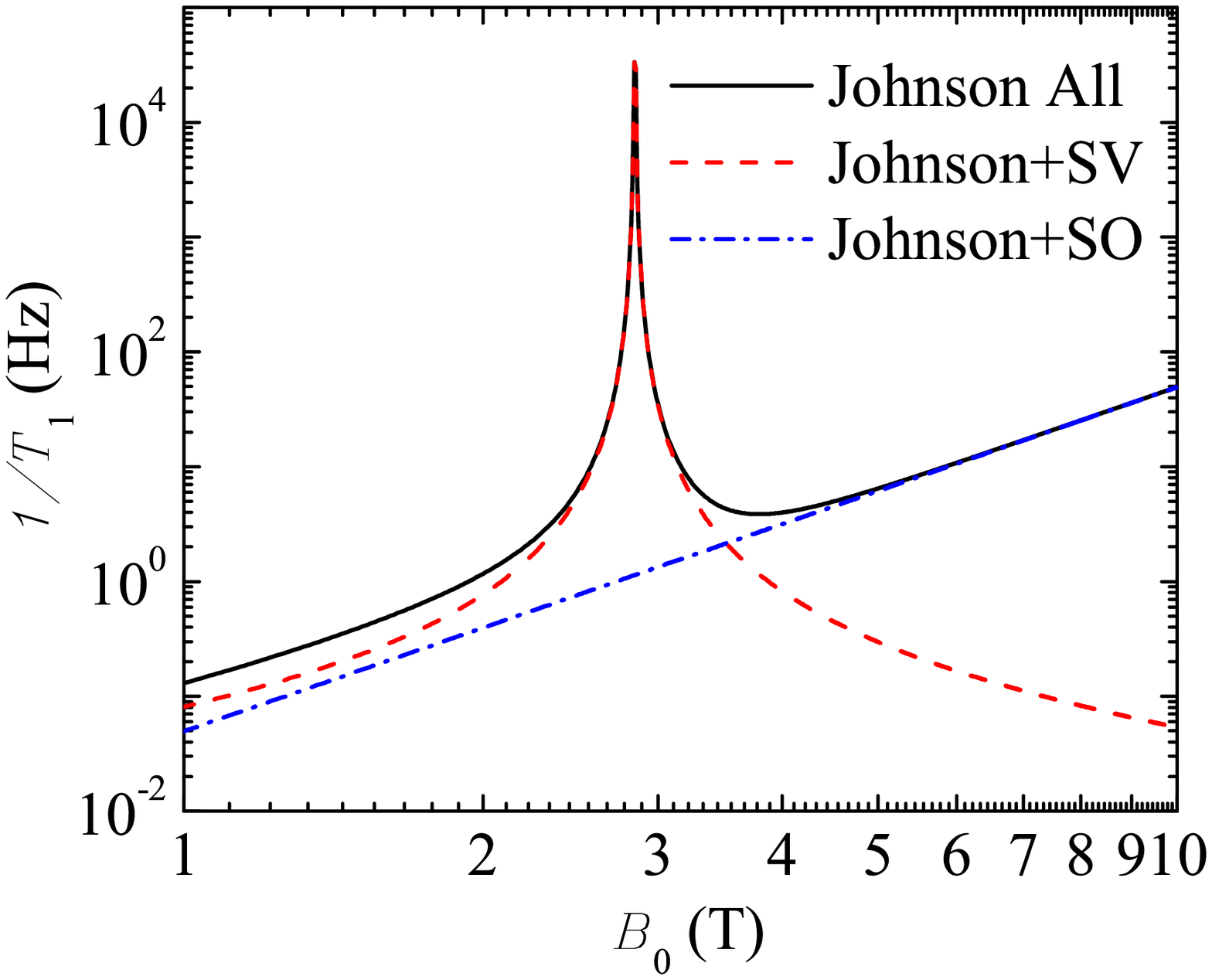}
\caption{Spin relaxation rate due to Johnson noise through SV mixing (red dashed line) and intra-valley SO mixing (blue dash-dotted line) as a function of in-plane magnetic field.} \label{Fig_Johnson}
\end{figure}

Figure~\ref{Fig_Johnson} shows the spin relaxation rates $\Gamma_\mathrm{SV}$ through SV mixing (red dashed line), $\Gamma_\mathrm{SO}$ through SO mixing (blue dash-dotted line), and the total spin relaxation rate $\Gamma_\mathrm{SV}+\Gamma_\mathrm{SO}$ (black solid line) as a function of the applied magnetic field $B_0$ due to Johnson noise. As shown in Fig.~\ref{Fig_Johnson}, the relaxation rate through the intra-valley SO mixing is dominant in the high-field regime, showing a $B_{0}^3$ dependence. The relaxation due to SV mixing peaks at the degenerate point ($g\mu_B B_0=E_\mathrm{VS}$), and it dominates in the low-magnetic-field regime due to a linear $\omega_Z$ dependence. The relaxation time due to the Johnson noise is about $10$ s when $B_0 = 1$ T, and about 0.01 s when $B_0 = 10$ T.





\subsection{Phonon noise}

\begin{figure}[tb]
\includegraphics[scale=0.44]{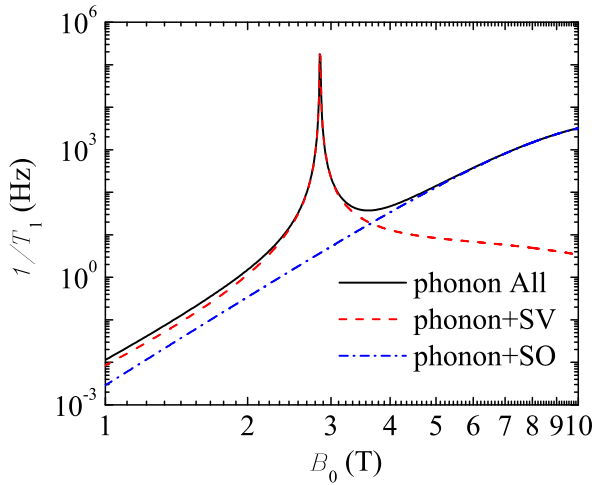}
\caption{Spin relaxation rate due to the deformation phonon noise through the SV mixing (red dashed line) and SO mixing (blue dash-dotted line) as a function of in-plane magnetic field.}\label{Fig_phonon}
\end{figure}

Phonon noise is the most studied spin relaxation source, and it is usually the dominant source of spin relaxation in the strong-magnetic-field regime because of the higher phonon density of states at high frequency.\cite{Khaetskii2001, Golovach2004, Amasha2008, Yang2013, Tahan2014, Huang2014} Although results for spin relaxation due to SV mixing and phonon noise have been obtained in Ref.~\onlinecite{Yang2013}, we include this spin relaxation channel here for completeness. Furthermore, a unified treatment is given here for both phonon and Johnson noise, and the phonon bottleneck effect is taken into account in a simplified manner.\cite{Tahan2014}

To obtain the results for phonon noise, we need the correlation of the electric field $E(\vec{r})=\vec{\nabla}U_\mathrm{ph}(\vec{r})/e$, which can be derived based on the electron-phonon interaction potential $U_\mathrm{ph}(\vec{r})$,\cite{Golovach2004, Tahan2014}
\begin{equation}
U_\mathrm{ph}(\vec{r},t) = \sum_{\boldsymbol{q}j} \frac{f(q_z) e^{i\vec{q}_\parallel\cdot\vec{r}}}{\sqrt{2\rho_c\omega_{qj}/\hbar}} (-iq\Xi_{\boldsymbol{q}j}) (b_{-\boldsymbol{q}j}^\dag + b_{\boldsymbol{q}j}), \label{Uph}
\end{equation}
where $b_{\boldsymbol{q}j}^{\dag }$ ($b_{\boldsymbol{q}j}$ ) creates (annihilates) an acoustic phonon with wave vector $\boldsymbol{q} = (\boldsymbol{q}_{\parallel }, q_{z})$, branch index $j$, and dispersion $\omega _{{q}j}$; $\rho _{c}$ is the sample density (volume is set to unity). The factor $f(q_{z})$ equals unity for $|q_{z}|\ll d^{-1}$ and vanishes for $|q_{z}|\gg d^{-1}$, where $d$ is the characteristic size of the quantum well along the $z$ axis. Here we consider the deformation potential electron-phonon interaction, with $\Xi_{\boldsymbol{q}j}$ being the deformation potential constants (piezo-electric interaction vanishes in Si due to the non-polar nature of the lattice). In Si, the deformation potential strength for different branches is $\Xi_{1} = \Xi_d + \Xi_u \cos^2 \theta$ (LA), $\Xi_{2}= 0$ (TA) and $\Xi_{3} = \Xi_u \cos \theta \sin\theta$ (TA), where $\Xi_d$, and $\Xi_u$ are the dilation and uniaxial shear deformation potential constants.\cite{Yu2010}


To calculate spin relaxation due to the phonon noise, we first need to obtain the phonon correlation functions, which are discussed in detail in
Appendix~\ref{append::phonon correlation}.  Substituting the correlation functions into Eq.~(\ref{Gamma_SV}), we find that the dependence of $\Gamma_\mathrm{SV}$ on the applied magnetic field is determined by the factor $\omega_{Z}^{5} F_\mathrm{SV}(\omega_Z)$. $\Gamma_\mathrm{SO}$, on the other hand, is proportional to $\omega _{Z}^{7}$. Both rates are proportional to the deformation potential strength $\Xi_j$ and inversely proportional to the seventh power of phonon velocity $v_j$.

Figure~\ref{Fig_phonon} shows the spin relaxation rates $\Gamma_\mathrm{SV}$ through SV mixing (red dashed line), $\Gamma_\mathrm{SO}$ through SO mixing (blue dash-dotted line), and the total spin relaxation $\Gamma_\mathrm{SV}+\Gamma_\mathrm{SO}$ (black solid line) as a function of the applied magnetic field $B_0$ due to phonon noise. The parameters are $\rho_c = 2200$ kg/m$^3$, $v_1 = 5900$ m/s, $v_2 =v_3= 3750$ m/s (data for SiO$_2$), $\Xi_d=5$ eV, $\Xi_u=8.77$ eV, $T=0.15$ K, and the other parameters are the same as before. Similar to Johnson noise, the relaxation through the SV mixing dominates in the low-magnetic-field regime, and it peaks at the degeneracy point. The relaxation rate through the intra-valley SO mixing is dominant in the high-magnetic-field regime, which shows a $B_{0}^7$ dependence before the phonon bottleneck takes effect and the curves bend downward from the $B_{0}^7$ line. \cite{Golovach2004, Tahan2014, Huang2014} The phonon bottleneck effect is due to the averaging of electron-phonon interaction matrix element for high-frequency phonons. This reduction in the effective coupling strength causes the spin relaxation rate to decrease from the $B_0^7$ curve in Fig.~\ref{Fig_phonon}, and it could even lead to a suppression of spin relaxation,\cite{Huang2014} as has been observed experimentally for a spin singlet-triplet qubit.\cite{Meunier2007}  Quantitatively, the relaxation time due to phonon noise is $\sim 100$ s in a 1 Tesla field, and $\sim 0.1$ ms in a 10 Tesla field.

\subsection{Comparison of Johnson and phonon noises}

In this section, we compare the magnetic-field dependence of the spin relaxation rate for Johnson noise and phonon noise. The effects of other noises, such as $1/f$ noise, are relatively small, as shown in Appendix \ref{append::1fnoise}. Since the magnetic-field dependence of spin relaxation is different for different noises, the dominant source of relaxation could be different in different regimes.

\begin{figure}[]
\includegraphics[scale=0.44]{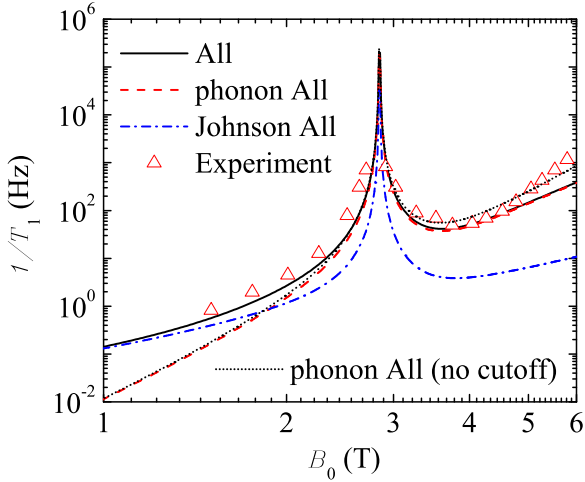}
\caption{Spin relaxation rate as a function of magnetic field in a Si QD with valley splitting $E_{vs}=0.33$ meV for phonon noise (red dashed line) and Johnson noises (blue dash-dotted line). The total spin relaxation is plotted as a black solid line, and the experimental results (red dots) are from Ref. \onlinecite{Yang2013}. For comparison, the result (black dotted thin line) of phonon-induced relaxation without considering the phonon bottleneck effect (without the cutoff function) is also presented, which reproduces the original fitting in Ref.~\onlinecite{Yang2013}.
} \label{Gamma_comp}
\end{figure}

Figure~\ref{Gamma_comp} shows the spin relaxation due to phonon noise (red dashed line) and Johnson noise (blue dash-dotted line) as a function of the applied magnetic field with a valley splitting $E_\mathrm{VS}=0.33$ meV. The other parameters are the same as in the previous two subsections, namely a QD confinement of $\hbar\omega_d=8$ meV, the dipole matrix elements $r_x^{-+} = r_y^{-+} = r_z^{-+} = 1.1$ nm, the SO interaction strengths $\alpha_R=45$ m/s and $\alpha_D=0$ m/s, and the resistance for Johnson noise at $R=2$ k$\Omega$. The red triangles are experimental results from Ref.~\onlinecite{Yang2013}. For comparison, the phonon-induced relaxation rates (black dotted line) obtained without considering the phonon bottleneck effect (without the cutoff function) are also presented, which reproduces the original fitting in Ref.~\onlinecite{Yang2013}.  There are three interesting features to this figure: the spin hot spot, which we have discussed extensively in previous subsections, the high-field trend, and the low-field trend.  Below we examine the later two features in more detail.

Figure~\ref{Gamma_comp} shows that, at high B-field, spin relaxation due to phonon noise dominates over relaxation due to Johnson noise, as expected from the spectral densities of these two noises.  At the highest magnetic fields in the figure, the curve without phonon bottleneck effect looks more consistent with the experimental data. This is because we are using the parameters from Ref. \onlinecite{Yang2013} instead of refitting the parameters such as the SO coupling $\alpha_R$ and the dipole matrix element $r^{-+}$. We emphasize that the only fitting parameter in our case is the resistance $R$. If we want more consistent results with experimental data, one needs to (i) increase the spin-orbit coupling $\alpha_R$ to have faster spin relaxation $\Gamma_\mathrm{SO}$ due to spin-orbit mixing; (ii) reduce the dipole matrix element $r^{-+}$, so that the width of the spin relaxation peak, which is determined by $\Delta_{23}$, does not change; and (iii) increase the resistance $R$ to get the same magnitude of spin relaxation at low fields. Since a slight variation of these parameters does not have much of an impact on the understanding of the system, and the parameters differs for different materials, we prefer using the parameters given by Ref. \onlinecite{Yang2013}, and changing only the resistance of Johnson noise to make sure that the low-frequency regime is well understood.  We also note that the measured relaxation rate seems to increase faster at very high fields ($>$ 4 T) than both theoretical calculations, with or without the phonon bottleneck effect.\cite{Yang2013}  This discrepancy could be due to another level crossing (and the associated spin hot spot) at a higher field that is not taken into consideration in the current study, or a reflection of non-parabolic features of the QD confinement.


At low magnetic fields, the dominant spin relaxation channel crosses over from phonon noise to Johnson noise (around 2 T).  As discussed in Sec. \ref{Johnson}, the dominant relaxation mechanism at low magnetic field is due to Johnson noise and SV mixing. By considering the Johnson noise, the theoretical results of total spin relaxation (black solid line) are now more consistent with the experimental measurements in Ref.~\onlinecite{Yang2013}, where the relaxation rate at $B=1$ T is around 0.1 s$^{-1}$.

\begin{figure}[]
\includegraphics[scale=0.75]{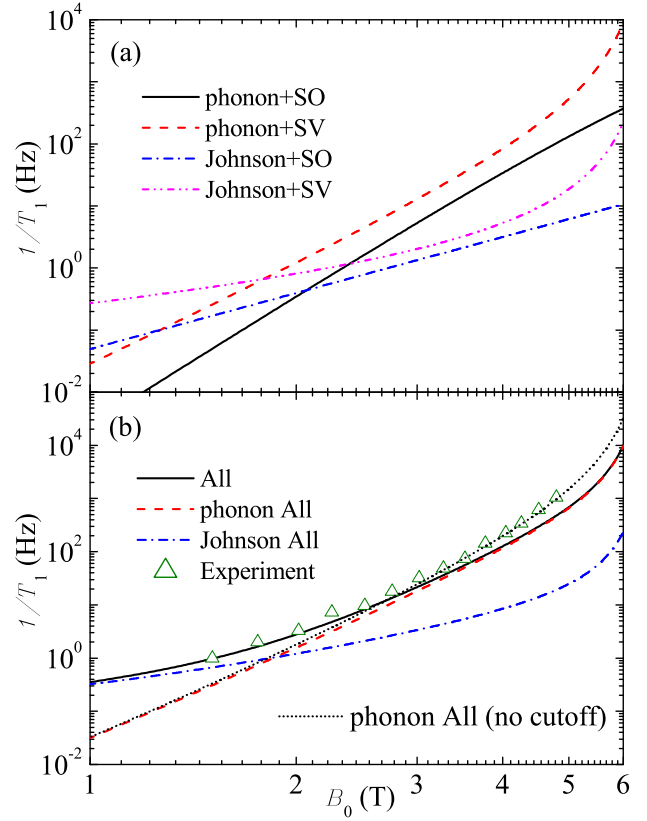}
\caption{Spin relaxation rate as a function of magnetic field in Si QD with valley splitting $E_{vs}=0.75$ meV for phonon and Johnson noises.  In panel (a), we compare the spin relaxation rates for phonon noise with SO mixing (black solid line) and SV mixing (red dashed line), and Johnson noise with SO mixing (blue dash-dotted line) and SV mixing (magenta dash-double-dotted line).  In panel (b), the total spin relaxation rate is plotted as a black solid line, and the total phonon contribution (red dashed line) and the total Johnson noise contribution (blue dot-dashed line) are also included.  The experimental results (green triangles) are taken from Ref.~\onlinecite{Yang2013}. For comparison, the result (black dotted thin line) of phonon-induced relaxation without considering the phonon bottleneck effect (i.e. no cutoff function) is also presented, which reproduces the original fitting in Ref.~\onlinecite{Yang2013}
} \label{Gamma_comp_Evs075}
\end{figure}

Figure~\ref{Gamma_comp_Evs075} shows the spin relaxation rate due to phonon noise and Johnson noise as a function of the applied magnetic field at a valley splitting of $E_{vs}=0.75$ meV.  The other parameters are the same as in Fig.~\ref{Gamma_comp}, except the dipole matrix elements are a bit larger at $r_x^{-+} = r_y^{-+} = r_z^{-+} = 1.7$ nm.\cite{Yang2013}   In essence, throughout the whole field range in this figure, the system is on the low-energy side of the degeneracy point or the spin hot spot.  As shown in Fig.~\ref{Gamma_comp_Evs075} (a), at higher magnetic fields, the dominant relaxation source is phonon noise and SV mixing. For lower fields, the dominant relaxation channel changes over to Johnson noise and SV mixing. Figure~\ref{Gamma_comp_Evs075} (b) shows that, similar to Fig.~\ref{Gamma_comp}, after including the effects of Johnson noise, the theoretical results of total spin relaxation (black solid line) are now more consistent with the experimental measurements (green triangles) at lower magnetic fields,\cite{Yang2013} where the relaxation rate at $B=1$ T is around 0.3 s$^{-1}$.

Figure~\ref{Gamma_comp_Evs075} is essentially the low-energy side of Fig.~\ref{Gamma_comp}, with a shift in the peak position and a slight increase in the peak width.  The enlarged plot does reveal more clearly one important fact: with the given Si parameters the phonons provide a more important relaxation channel compared to Johnson noise at the spin hot spot.  The transition of the dominant relaxation channel happens at a field significantly below the degeneracy point,
at just below 2 T. Again the no-cut-off results seem to fit the experimental data better than results with the phonon bottleneck effect. This is due to our choice of parameters $\alpha_D$ and $r^{-+}$, which are taken directly from Ref.~\onlinecite{Yang2013}. Since a slight variation of these parameters does not change our understanding of spin dynamics, we use the values of these parameters from Ref.~\onlinecite{Yang2013}, and we change only the resistance of Johnson noise to make sure that the data fit in the low frequency regime is optimized.


\section{Conclusion}
\label{sec:conclusion}

In conclusion, we have studied spin relaxation of an electron in a Si QD with valley splitting.  In particular, we have clarified how the presence of conduction-band valleys influences spin relaxation. By considering both spin-valley mixing and intra-valley spin-orbit mixing in a Si QD, we find that spin relaxation due to Johnson noise is the dominant spin relaxation channel (as compared to phonons and other electrical noises) when the Zeeman splitting is much smaller than the valley splitting.

In our calculations, we have included both Johnson and phonon noises, and we incorporated both spin-valley and intra-valley spin-orbit mixings. For the various field regimes as compared with valley splitting we find the following.  In the low-field regime, when Zeeman splitting $E_Z$ is much smaller than the valley splitting $E_\mathrm{VS}$, Johnson noise together with spin-valley mixing leads to the fastest spin relaxation because of a weaker field-dependence.  As the magnetic field increases and the Zeeman splitting approaches the valley splitting, $E_Z \sim E_\mathrm{VS}$, spin-valley mixing together with both phonon noise and Johnson noise produces a sharp peak in the spin relaxation rate, though for Si with the parameters from experiments, phonon noise is now the most important source of spin relaxation (while Johnson noise also contributes significantly). When the applied field increases further, $E_Z > E_\mathrm{VS}$, the intra-valley spin-orbit mixing gradually becomes the dominant spin relaxation mechanism because of its stronger dependence on the external field, which is consistent with the existing literature. Using parameters obtained from an experimental measurement \cite{Yang2013}, and a single fitting parameter of low-temperature circuit resistance, we obtain numerical results that fit the measurements well in the whole range of applied magnetic field.

We acknowledge the support of U.S. ARO (W911NF0910393) and NSF PIF (PHY-1104672). We also acknowledge useful discussions with Andrew Dzurak, Andrea Morello, Jason Petta, and Charles Tahan. XH would also like to acknowledge the hospitality of Kavli Institute of Theoretical Physics China, where part of this work was completed.

\appendix

\section{Effects of the magnetic field orientation}
\label{append:B-orientation}

The spin relaxation mechanism we study in this paper involves the spin-orbit interaction.  When both Dresselhaus and Rashba SO coupling are present in a system, such as in a Si heterostructure, the orientation of the applied magnetic field plays an important role in determining the amount of transverse magnetic noise and thus the relaxation rate.  Here we discuss this field orientation dependence in detail.

Consider a magnetic field in an arbitrary direction, $\boldsymbol{B}_{0}=B_{0}\left(\sin\theta\cos \phi ,\sin\theta\sin \phi ,\cos\theta\right)$, where $\theta $ and $\phi $ are the polar and azimuthal angles of the magnetic field in the ($xyz$) coordinate system. By using the relationship $\vec{p}^{-+}=\bra{-}\vec{p}\ket{+}=im^*E_\mathrm{VS} \vec{r}^{-+}/\hbar$, the spin-valley mixing matrix element $\Delta_{23}$ can be expressed in terms of the electric dipole matrix element,\cite{Tahan2014}
\beqa
\Delta_{23} &=& \frac{2im^*E_\mathrm{VS}}{\hbar} \left[\alpha_-r _{y}^{-+}\sigma _{x}^{\uparrow\downarrow} + \alpha_+ r_{x}^{-+} \sigma_{y}^{\uparrow\downarrow}\right],
\eeqa
where $\vec{\sigma}^{\uparrow\downarrow}=\bra{\uparrow}\vec{\sigma}\ket{\downarrow}$ is the spin flip matrix elements, and $\alpha_\pm$ are the spin-orbit coupling constants.

In order to calculate the spin flip matrix elements $\vec{\sigma}^{\uparrow\downarrow}$, it is convenient for us to express the spin state $\ket{\psi_\mu}$ ($\ket{\psi_\mu}$=$\ket{\uparrow}$ or $\ket{\downarrow}$), which are the eigenfunctions of $\sigma_{z^{\prime}}$ ($z^\prime$ axis along the magnetic field), in terms of the eigenstates $\ket{\chi_m}$ of $\sigma_z$: $\ket{\psi_\mu}=\sum_{m=\pm1/2}D^{(1/2)*}(\phi,\theta,0)\ket{\chi_m}$, where $D^{(1/2)}$ is the finite rotation matrix,\cite{Landau1977}
\beqa
\ket{\psi_\uparrow}&=&e^{-i\phi/2}\cos\theta/2\ket{\chi_\uparrow}+e^{i\phi/2}\sin\theta/2\ket{\chi_\downarrow},\\
\ket{\psi_\downarrow}&=&-e^{-i\phi/2}\sin\theta/2\ket{\chi_\uparrow}+e^{i\phi/2}\cos\theta/2\ket{\chi_\downarrow}.
\eeqa
Therefore, spin-flip matrix elements are $\sigma _{x}^{\uparrow\downarrow}=\cos\theta\cos\phi+i\sin\phi$ and $\sigma _{y}^{\uparrow\downarrow}=\cos\theta\sin\phi-i\cos\phi$, and
the square of the magnitude of the SV mixing matrix element is
\beqa
\lefteqn{|\Delta_{23}|^2 = (2m^*E_\mathrm{VS}/\hbar)^2\left\{\alpha_-^2|r _{y}^{-+}|^2(\cos^2\theta\cos^2\phi+\sin^2\phi)\right.}\nonumber\\
&+&\alpha_+^2|r _{x}^{-+}|^2(\cos^2\theta\sin^2\phi+\cos^2\phi)+2\alpha_-\alpha_+\mathrm{Re}[r _{y}^{-+}r _{x}^{+-}\nonumber\\
&\times&\left.(-\sin^2\theta\cos\phi\sin\phi+i\cos\theta)]\right\}.
\eeqa

When $\boldsymbol{B}_{0}$ is along the z-direction ($\theta=0$, $\phi=0$),
\beqa
\lefteqn{|\Delta_{23}|^2 = (2m^*E_\mathrm{VS}/\hbar)^2\left\{\alpha_-^2|r _{y}^{-+}|^2 \right.}\nonumber\\
&+&\left. \alpha_+^2|r _{x}^{-+}|^2 - 2\alpha_- \alpha_+ \mathrm{Im}[r _{y}^{-+}r _{x}^{+-}]\right\}.
\eeqa
When $\boldsymbol{B}_{0}$ is in the plane of 2DEG ($\theta=\pi/2$),
\beqa
\lefteqn{|\Delta_{23}|^2 = (2m^*E_\mathrm{VS}/\hbar)^2\left\{\alpha_-^2|r _{y}^{-+}|^2\sin^2\phi\right.}\nonumber\\
&+&\left.\alpha_+^2|r _{x}^{-+}|^2\cos^2\phi - \alpha_-\alpha_+\mathrm{Re}[r _{y}^{-+}r _{x}^{+-}\sin2\phi]\right\}.
\eeqa
Therefore, the magnetic field orientation dependence of $\Delta$ (or $\Gamma_\mathrm{SV}$) depends on the values of $\alpha_-$, $\alpha_+$, $r_{y}^{-+}$ and $r_{x}^{+-}$, which is material- and device-specific. In particular, if the magnetic field is along the [110] crystal axis $\phi=0$, as is the case in Ref.~\onlinecite{Yang2013}, $\sigma_{x}^{\uparrow\downarrow}=0$, $\sigma _{y}^{\uparrow\downarrow}=-i$, and $\Delta_{23} = 2m^*E_\mathrm{VS}\alpha_+ r_{x}^{-+}/\hbar$.  In our calculation, we used $\Delta_{23} = m^*E_\mathrm{VS}\alpha_R r _{x}^{-+}/2/\hbar$ to reproduce the results of Ref.~\onlinecite{Yang2013}.

\section{$1/f$ charge noise and spin relaxation}
\label{append::1fnoise}

The $1/f$ charge noise is quite common in semiconductor devices, and is often believed to be an important decoherence source for charge qubits.  Here we explore how much it affects a spin qubit.

The $1/f$ charge noise is often measured via the fluctuations it causes in the energy levels in a quantum dot or a quantum point contact (QPC). \cite{Jung2004, Buizert2008, Hitachi2013, Takeda2013} Consider the current through a QPC connected to two leads.  The current is sensitively dependent on the gate voltage applied to the QPC. By measuring the electric current fluctuations, the overall effect of the $1/f$ charge noise on the QPC can be measured.  Normally, such an experiment has a finite frequency range, e.g. from a few Hz to hundreds of Hz. The measured energy level fluctuations actually depend on the frequency range of the measurement, and are thus dependent on the specific experiment. Thus here we first try to extract a quantity that is independent of the frequency range in these experiments.

We assume the current fluctuation spectral density due to the $1/f$ charge noise in a QPC to be $S_I(\omega)=A_I/\omega$.  An integration of the spectrum yields
\beq
\int_{\omega_0}^{\omega_c} d \omega S_I(\omega)=A_I \ln \frac{\omega_c}{\omega_0} \,.
\eeq
Phenomenologically, the current fluctuation can be represented by an effective gate voltage fluctuation, \cite{Jung2004, Buizert2008, Hitachi2013, Takeda2013}
\beq
\Delta V_{EG}=\sqrt{2\int_{\omega_0}^{\omega_c}d \omega S_I(\omega)}\bigg/\frac{d I_{QPC}}{dV_G},
\eeq
where $\omega_0$ and $\omega_c$ are the lower and upper cutoff frequency (response frequency) in the experiment. $dI_{QPC}/dV_{G}$ is the effective differential conductivity, which represents the variation of the electric current through QPC due to the gate voltage difference. Therefore, the quantity $\Delta V_{EG}$ represents the effective gate voltage fluctuation due to charge noise in the system. In order to get the effective electric field on the electron in the QD, we should also consider the screening effect of the gate voltage.

The quantity $\Delta V_{EG}$ defined here is dependent on the frequency range of the measurement in the experiments,
\beq
\Delta V_{EG}=\sqrt{2A_I}\ln\left(\frac{\omega_c}{2\omega_0}\right)\bigg/\frac{d I_{QPC}}{dV_G}.
\eeq
We define a quantity $\Delta \tilde{V}_{EG}=\Delta V_{EG}/(\sqrt{2}\ln(\omega_c/2\omega_0))$ as the effective gate voltage fluctuation, which is independent of the frequency range. Take Ref.~\onlinecite{Takeda2013} as an example for the $1/f$ charge noise in Si/SiGe, where $\Delta V_{EG}=0.1$ meV, $\omega_0=0.01$ Hz, $\omega_c=49$ Hz, and $\sqrt{2}\ln(\omega_c/2\omega_0)=11.03$. Therefore, the effective gate voltage fluctuation due to charge noise is $\Delta \tilde{V}_{EG}\approx 10$ $\mu$eV. Due to the screening of the gate voltage, the effective voltage fluctuation sensed by the electron in the QD is around $1$ $\mu$eV.

\begin{figure}[tb]
\includegraphics[scale=0.44]{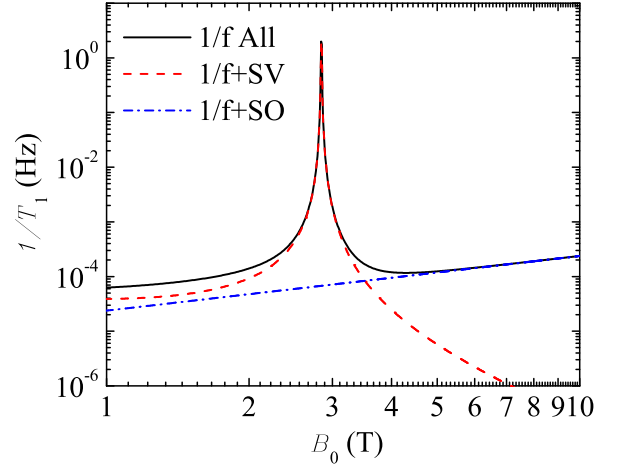}
\caption{Spin relaxation rate due to $1/f$ charge noise through SV mixing (red dashed) and SO mixing (blue dash-dotted) as a function of in-plane magnetic field.}\label{Fig_1f}
\end{figure}

With the knowledge of the magnitude of $1/f$ charge noise, we can calculate the corresponding spin relaxation. The spin relaxation due to the SV mixing and $1/f$ charge noise is given by
\beq
\Gamma_\mathrm{SV}=\frac{2\pi}{\hbar^2}\sum_i\left|r^{-+}_i\right|^2 e^2 A \omega _{Z}^{-1}F_\mathrm{SV}(\omega_Z),
\eeq
where $r^{-+}_i$ are the transition matrix elements between the two lowest valley states, $A$ is the charge noise amplitude, and $\omega_Z$ is the Zeeman frequency. The dependence of $1/T_{1}$ on the applied magnetic field is $1/T_{1}\propto B_{0}^{-1}F_\mathrm{SV}(g\mu_BB_0/\hbar)$, and the function $F_\mathrm{SV}(\omega_Z)$ is given by Eq.~(\ref{FSV}).


Figure~\ref{Fig_1f} shows the spin relaxation rates $\Gamma_\mathrm{SV}$ through SV mixing (red dashed line), $\Gamma_\mathrm{SO}$ through SO mixing (blue dash-dotted line) and the total spin relaxation $\Gamma_\mathrm{SV}+\Gamma_\mathrm{SO}$ (black solid line) as a function of the applied magnetic field $B_0$ due to $1/f$ charge noise. The results of the spin relaxation rate $\Gamma_\mathrm{SO}$ due to charge noise and intra-valley SO mixing is from Ref. \cite{Huang2014}. As shown in the figure, the relaxation through the mechanism of SV mixing dominates in the low magnetic field regime, and it peaks at the degenerate point ($g\mu_B B_0=E_\mathrm{VS}$). The relaxation rate through the intra-valley SO mixing is dominating in the high magnetic field regime. The relaxation time due to the $1/f$ charge noise is about $10^4$ s for a Si QD, when the Zeeman energy is away from the valley splitting.


\section{Spectrum of Phonon Noise}
\label{append::phonon correlation}

The electron phonon interaction $U_\mathrm{ph}(\vec{r})$ is given by Eq. (\ref{Uph}). In the interaction picture, the electron phonon interaction acquires a time dependence, with $b_{\boldsymbol{q},j}(t)=b_{\boldsymbol{q},j}e^{-i \omega_{\boldsymbol{q},j}t}$ and $b_{\boldsymbol{q},j}^{\dag}(t) = b_{\boldsymbol{q},j}^{\dag}e^{i \omega_{\boldsymbol{q},j}t}$.
The correlation of the electric force due to phonons, $-eE(\vec{r})=-\vec{\nabla}U_\mathrm{ph}(\vec{r})$, is thus given by (x component),
\begin{eqnarray}
\lefteqn{e^2\left\langle E_x(0)E_x(t)\right\rangle = \sum_{\boldsymbol{q}j}\frac{\left|f(q_z)\right|^2} {{2\rho_c\omega_{qj}/\hbar}} q_x^2e^{i\vec{q}_\parallel\cdot\vec{r}} }\nonumber\\
&&\times  \left|q\Xi_{\boldsymbol{q}j}\right|^2(b_{\boldsymbol{q}j}b_{\boldsymbol{q}j}^\dag e^{i\omega_{qj}t}+b_{-\boldsymbol{q}j}^\dag b_{-\boldsymbol{q}j}e^{-i\omega_{qj}t}). \label{ExEx}
\end{eqnarray}

We consider the adiabatic condition, where the energy scale of the noise is much less than the dot confinement energy $E_d=\hbar\omega_d$ and the valley splitting, so that the electron orbital state stays in the instantaneous ground state
$
\psi (\vec{r})=\exp\left( -(\vec{r}-\vec{r}_0)^2/2\lambda ^{2}\right) /\lambda \sqrt{\pi },
$
where $\lambda ^{-2}=\hbar ^{-1}\sqrt{(m^{\ast }\omega _{d})^{2}+(eB_{z}/2c)^{2}}$ is the effective radius. Then, we simplify the exponential terms $e^{i\vec{q}_\parallel\cdot\vec{r}}$ by its mean-field value
$e^{-q_\parallel^2\lambda^2/4}$.

The summation in Eq.~(\ref{ExEx}) for all possible $\boldsymbol{q}$ in the momentum space can be expressed as integrals
\begin{eqnarray}
\sum_{j}\int d\omega d\theta d\varphi D_{j}(\omega,\theta)g_j(\omega,\theta,\varphi), \label{sumq}
\end{eqnarray}
where $D_{j}(\omega,\theta)= \frac{1}{(2\pi)^3}\frac{\omega^{2}}{v_j^3}\sin \theta$ is the density of states for phonons, and
\beqa
\lefteqn{g_j(\omega,\theta,\varphi,t)=\frac{\hbar\bar{\Xi}_{j}^2}{2\rho_cv_j^4}\sin^2\theta\cos^2\varphi }\nonumber\\
&&\times\omega^3\left[(N_{\omega}+1)e^{i\omega t}+N_{\omega}e^{-i\omega t}\right] f_j(\omega,\theta). \label{fj}
\eeqa
In Eq.~(\ref{fj}), $N_{\omega}=(\exp(\hbar\omega/k_BT)-1)^{-1}$ is the phonon excitation number and the cutoff function $f_j(\omega,\theta) = \left|f\left({\omega}\cos\theta/{v_j}\right)\right|^2e^{-\omega^2\lambda^2\sin^2\theta/2v_j^2}$ is due to
the suppression of the matrix element for the electron-phonon interaction in a large QD.\cite{Huang2014}

The spectrum of the phonon noise in the $x$-direction is therefore ($\int_0^{2\pi} d\varphi \cos^2\varphi=\pi$)
\beqa
\lefteqn{S_{xx}^E(\omega)=\mathrm{Re}\frac{1}{2\pi}\int_{-\infty}^{\infty}dt\, \langle E_{x} E_{x}(t)\rangle \cos(\omega t)} \nonumber\\
&=& \sum_{j}\frac{\hbar\omega^5(2N_{\omega}+1)}{16\pi^2e^2\rho_cv_j^7}\int_0^{\pi/2} d\theta \bar{\Xi}_{j\theta}^2\sin^3\theta f_j(\omega,\theta).
\nonumber
\eeqa
Similarly, $S_{yy}^E(\omega)=S_{xx}^E(\omega)$ and
\beq
S_{zz}^E(\omega)=\sum_{j}\frac{\hbar\omega^5(2N_{\omega}+1)}{8\pi^2e^2\rho_cv_j^7}\int_0^{\pi/2} d\theta \bar{\Xi}_{j\theta}^2\sin\theta\cos^2\theta f_j(\omega,\theta).
\nonumber
\eeq
If the dipole approximation $e^{i\vec{q}_\parallel\cdot\vec{r}}\approx 1+ i\vec{q}_\parallel\cdot\vec{r}$ is employed (for most spin qubit applications, the dipole approximation should be valid), so that $f_j(\omega,\theta)=1$, the relaxation rate would have taken the form given in Ref.~\onlinecite{Tahan2014}.  Furthermore, the temperature $T$ of the lattice vibration is normally very low ($T<1$ K), so that $2N_{\omega}+1=\coth(\hbar\omega/2k_{B}T)\approx 1$, in which case the spectrum of phonon noise shows a nice $\omega^5$ dependence.

\end{document}